\documentclass[12pt,twocolumn]{iopart}

\expandafter\let\csname equation*\endcsname\relax

\expandafter\let\csname endequation*\endcsname\relax

\usepackage{amsmath}
\usepackage{graphicx}% Include figure files
\usepackage{dcolumn}% Align table columns on decimal point
\usepackage{braket}
\usepackage{tcolorbox}

\def\bcen{\begin{center}}
\def\ecen{\end{center}}
\usepackage{epstopdf} 
\usepackage{tabularx}
\begin{document}

\title{Emergence and Persistence of Maximal Atom-Field Entanglement in the Jaynes-Cummings model}

\author{Siva Prasanna Karri}
\address{Department of Physics, SRM University-AP, Amaravati, Andhra Pradesh 522240, India}
\author{Mounika Sri Gudipati}
\address{Department of Physics, SRM University-AP, Amaravati, Andhra Pradesh 522240, India}
\author{Nilakantha Meher}
\address{Department of Physics, SRM University-AP, Amaravati, Andhra Pradesh 522240, India}
\ead{nilakantha.meher6@gmail.com}
\vspace{10pt}

\begin{abstract}
We investigate the emergence and persistence of maximal entanglement between a two-level atom and a single-mode quantized electromagnetic field inside a cavity, whose dynamics are governed by the Jaynes-Cummings Hamiltonian. Using the von Neumann entropy as a measure of entanglement, we establish a direct connection between entanglement, atomic coherence, and atomic population inversion. This establishment shows that maximal atom–field entanglement is achieved when both the atomic population inversion and the $l_1$-norm of atomic coherence simultaneously vanish. By considering the atom initially in its excited state, we further demonstrate that the persistence of maximal atom–field entanglement strongly depends on the initial state of the field. For field states whose photon-number distributions have no two consecutive occupied photon-number levels, the local atomic coherence remains zero throughout the evolution.  Consequently, complete collapse of the atomic population inversion for an extended period leads to persistent maximal atom–field entanglement. In contrast, for other field states, the interaction generates local atomic coherence, resulting in transient maximal entanglement. Our results demonstrate how the choice of the initial field state controls the temporal characteristics of entanglement in the Jaynes–Cummings system. The generation and persistence of maximal atom–field entanglement in such systems could be valuable for applications in quantum information processing. 
\end{abstract}

\vspace{2pc}
\submitto{\PS}
\maketitle

\section{Introduction}
Quantum entanglement is an interesting phenomenon in which two or more quantum systems become correlated so that the quantum state of each system cannot be described independently \cite{einstein1935can}. Instead, the composite system must be described by a single joint quantum state, even if the subsystems are separated by a large distance \cite{bell1964einstein}. There are several applications of quantum entanglement, including quantum teleportation \cite{bennett1993teleporting}, quantum dense coding \cite{PhysRevLett.69.2881}, and quantum cryptography \cite{bennett2014quantum, bennett1992quantum}. Therefore, the generation and quantification of entanglement have attracted considerable attention in recent years.  Several theoretical proposals have been put forward for the generation of entangled states in various physical systems, including cavity-QED systems \cite{haroche1989cavity,RevModPhys.85.1083,RevModPhys.87.1379,meher2022review}, superconducting systems  \cite{blais2004cavity,Wallraff2004,blais2021circuit}, and optomechanical systems \cite{aspelmeyer2014cavity}.  In cavity QED systems, schemes for high-fidelity generation of entanglement between atoms \cite{Ritter2012,RevModPhys.73.565,PhysRevLett.79.1,PhysRevA.50.R2799}, between cavities \cite{Liew2013,meher2017duality,PhysRevA.84.052302,PhysRevA.83.062310,IKRAM2000417}, and between atoms and cavities \cite{RevModPhys.73.565,PhysRevLett.90.027903,ArevaloAguilar2006} have been proposed, and a few of them have been realized experimentally \cite{Axline2018,PhysRevA.64.050301,PhysRevA.50.R895,PhysRevLett.106.060401}. However, in addition to the generation of entangled states, understanding the mechanisms responsible for the emergence of entanglement and its persistence for a longer period is equally important \cite{gabbassov2025adiabatic, amrita2021analysis, xie2019dynamics, pandit2018effects}.

In this study, we investigate the emergence of entanglement in an atom-cavity system governed by the Jaynes-Cummings interaction and its persistence during time evolution. The Jaynes-Cummings model describes the interaction between a single-mode quantized field and a two-level atom \cite{jaynes1963comparison}. Since its introduction, it has become a standard model for understanding the complex interaction between light and matter in quantum optics \cite{ShoreKnight1993, larson2021jaynes}. It explains many experimentally observed phenomena,  such as spontaneous emission \cite{purcell1995spontaneous}, vacuum-Rabi splitting \cite{thompson1992observation,PhysRevLett.93.233603}, photon blockade in a cavity \cite{birnbaum2005photon}, the collapse and revival of atomic population inversion \cite{jaynes1963comparison,ShoreKnight1993,cummings1965stimulated,gerry2023introductory}, etc., that cannot be explained if we assume the electromagnetic field to be a classical wave. Moreover, the model has been investigated for generating various nonclassical states of light, including Fock states \cite{Premaratne2017,PhysRevLett.125.093603,PhysRevA.50.3340,Varcoe2000,PhysRevLett.88.143601}, squeezed states \cite{ourjoumtsev2011observation,ZHENG2007460}, Schrödinger-cat states \cite{PhysRevA.45.5193}, and number-state filtered coherent states \cite{Meher2018,Meher2019}. 

Here, in an atom-field system whose dynamics is governed by the Jaynes-Cummings Hamiltonian \cite{jaynes1963comparison},  we find a direct connection between the dynamics of atomic observables and entanglement generated through the atom–field interaction. This connection shows that the emergence of maximal atom–field entanglement can be understood in terms of the simultaneous vanishing of the atomic population inversion and the local coherence of the atom, specifically the $l_1$-norm of coherence \cite{Baumgratz2014Quantifying}.  Starting with the atom in its excited state, for which the $l_1$-norm of coherence of the atom is zero, we consider various initial field states for studying the temporal characteristics of entanglement and demonstrate that the persistence of maximal entanglement for a longer duration depends strongly on the initial state of the field. For the field states having a photon-number distribution with no two consecutive photon-number states occupied, the $l_1$-norm of coherence remains zero throughout the evolution. In such cases, the complete collapse of the atomic population inversion directly leads to maximal atom–field entanglement, which persists over an extended period. In contrast, for other initial field states, the atomic coherence develops during the evolution, and the $l_1$-norm of coherence becomes non-zero. As a result, maximal entanglement is attained only for a short duration, when both the $l_1$-norm of coherence and atomic population inversion vanish.

\section{Physical System: The Jaynes-Cummings Model}
Consider a two-level atom interacting with a single-mode quantized electromagnetic field inside a cavity. The atom-field interaction can be described by the Jaynes-Cummings Hamiltonian, given by \cite{jaynes1963comparison,ShoreKnight1993,RevModPhys.85.1083,RevModPhys.87.1379,meher2022review}
\begin{align}\label{HJC}
H = \frac{1}{2}\hbar\omega_0 \sigma_z + \hbar\omega \left(a^\dagger a + \frac{1}{2}\right) + \hbar g \left( a^\dagger \sigma_- + a \sigma_+ \right),
\end{align}
where $\omega_0$ is the atomic transition frequency, $\omega$ is the cavity resonance frequency, $g$ is the atom–field coupling strength, $a(a^\dagger)$ is the photon annihilation (creation) operator for the field mode, $\sigma_z$ is the atomic energy operator and $\sigma_+ = \ket{e}\bra{g}, \, \sigma_- = \ket{g}\bra{e}$ are the atomic raising and lowering operators respectively. Under the resonance condition $\omega=\omega_0$, the energy eigenvalues of the Hamiltonian $H$ are $E_{\pm} = \hbar \omega (n+1) \pm \hbar g \sqrt{n+1}$ with the corresponding eigenvectors
$\ket{n_{\pm}} = \frac{1}{\sqrt{2}} \left( \ket{e,n} \pm \ket{g,n+1} \right).$

Let the atom initially be in its excited state  $\ket{\psi_{{a}}(0)} = \ket{e}$, and the field mode be in an arbitrary superposition of Fock states  $\ket{\psi_{{f}}(0)} = \sum_{n=0}^{\infty} C_n \ket{n}$, where $C_n=\langle n|\psi_{f}(0)\rangle$ is the probability amplitude for the $n$-photon state. These amplitudes can be calculated by expressing the initial field state in the Fock basis. We have provided the Fock basis representation of various field states in the \ref{AppendixCalculation}.

Under the evolution governed by the Hamiltonian $H$ with $\omega=\omega_0$, the evolved state of the atom-field system at time $t$ is \cite{gerry2023introductory,Sugumar2026}
\begin{align}\label{wft}
\ket{\Psi_{af}(t)} = \sum_{n=0}^{\infty} C_n e^{-i(n+1)\omega t}\left[ \cos\!\big({\Omega}_n\,t\big)\ket{e,n} - i \sin\!\big({\Omega}_n\,t\big)\ket{g,n+1}\right], 
\end{align}
where ${\Omega}_n=g\sqrt{n+1}$ is the Rabi frequency. For the non-resonant conditions, the evolved state of the atom-field system is given in the \ref{AppendixNonresonant}. Throughout this work, we consider the resonant condition unless the non-resonant condition is explicitly specified.

\subsection{Atomic coherence}
For a two-level atom, the density matrix in the basis $\{\ket{e},\ket{g}\}$ is of the form
\begin{align}\label{atomicdensitymatrixform}
\rho_a =\begin{bmatrix}\rho_{ee} & \rho_{eg} \\ \rho_{ge} & \rho_{gg}\end{bmatrix}. 
\end{align}
If $\rho_{eg}=\rho_{ge}=0$, then the state is maximally incoherent \cite{Baumgratz2014Quantifying}. Therefore, the off-diagonal elements of the density matrix represent the coherence between $\ket{e}$ and $\ket{g}$, which can be quantified by the $l_1$-norm of coherence. The $l_1$-norm of coherence of the quantum state $\rho_a$ is defined as \cite{Baumgratz2014Quantifying}
\begin{align}\label{l1normcoherence}
C_{l_1}=|\rho_{eg}|+|\rho_{ge}|=2|\rho_{eg}|.
\end{align}

The density matrix corresponding to the atom-field state is $\rho_{af}(t) = \ket{\Psi_{af}(t)}\!\bra{\Psi_{af}(t)}$. By taking the partial trace over the field degree of freedom, we get the reduced density matrix for the atom to be \cite{ShoreKnight1993}
\begin{align}\label{density matrix}
\rho_a(t) =\mathrm{Tr}_f \big[\rho_{af}(t) \big]=\begin{bmatrix}\rho_{ee}(t) & \rho_{eg}(t) \\ \rho_{ge}(t) & \rho_{gg}(t)\end{bmatrix}, 
\end{align}
where the matrix elements are
\begin{subequations}
\begin{align}
\rho_{ee}(t) &= \sum_{n=0}^{\infty} |C_n|^2 \cos^2\!\big({\Omega}_n\,t\big),\\
 \rho_{eg}(t) &= i e^{-i\omega t} \sum_{n=0}^{\infty} C_{n+1} C_n^* 
\cos\!\big({\Omega}_{n+1}\,t\big)\,\sin\!\big({\Omega}_n\,t\big),\\
\rho_{ge}(t) &= -i e^{i\omega t} \sum_{n=0}^{\infty} C_n C_{n+1}^*
\cos\!\big({\Omega}_{n+1}\,t\big)\,\sin\!\big({\Omega}_n\,t\big),\\
\rho_{gg}(t) &= \sum_{n=0}^{\infty} |C_n|^2 \sin^2\!\big({\Omega}_n\,t\big).
\end{align}
\end{subequations}

Then, we find the $l_1$-norm of coherence of the atom to be
\begin{align}\label{atomiccoherence}
C_{l_1} (t)=2\left|\sum_{n=0}^{\infty} C_{n+1} C_n^* 
\cos\!\big({\Omega}_{n+1}\,t\big)\,\sin\!\big({\Omega}_n\,t\big)\right|.
\end{align}

\subsection{Atomic population inversion}
The atomic population inversion, defined as the difference between the probabilities of finding the atom in its excited state and ground state, can be calculated for the state given in Eq. \eqref{wft} as
\begin{align}\label{populationinversion}
 W(t)=\rho_{ee}(t)-\rho_{gg}(t) =\sum_{n=0}^\infty P_n \cos\left(2\Omega_n t\right),
\end{align}
where $P_n=|C_n|^2$ is the probability of finding $n$ photons in the field state $\ket{\psi_{f}}$. The atomic population inversion $W(t)=1$ when the atom is in the excited state, whereas $W(t)=-1$ when the atom is in the ground state. However, for some field states, $W(t) \approx 0$ for a longer period of time \cite{jaynes1963comparison,ShoreKnight1993,gea1990collapse,PhysRevA.45.8190}, which can be attributed to destructive interference between the cosine terms in Eq. \eqref{populationinversion}. This phenomenon is known as the collapse of atomic population inversion \cite{gerry2023introductory}.

\section{Emergence of Maximal Atom-field entanglement}
To investigate the emergence of maximal entanglement between the two-level atom and the cavity field, we consider the von Neumann entropy \cite{vonNeumann1927,von2018mathematical} to quantify entanglement in the pure state $|\Psi_{af}(t) \rangle$. For any bi-partite pure state, here the atom-field state $|\Psi_{af}(t) \rangle$, we have 
\begin{align}\label{vonNeumann}
S (t)= -\text{Tr}(\rho_a (t) \log_2 \rho_a (t)),
\end{align}
where $\rho_a (t)=\text{Tr}_f(|\Psi_{af}(t) \rangle \bra{|\Psi_{af}(t)})$ is the reduced density matrix of the atom given in Eq. \eqref{density matrix}. The quantity $S$ varies between 0 and 1. If $S=0$, then the composite state $|\Psi_{af}(t) \rangle$ is separable, indicating the absence of entanglement between the atom and the field. For a non-zero value of $S$, the state $|\Psi_{af} (t)\rangle$ is entangled, and for $S=1$, it is maximally entangled.   

Using Eq. (\ref{density matrix}), Eq. \eqref{vonNeumann} is simplified to
\begin{align}\label{vonNeumannJC}
S (t)= -\lambda_+ \log_2 \lambda_+ - \lambda_- \log_2 \lambda_-,
\end{align}
where $\lambda_{\pm}=\tfrac{1}{2}\Big(1 \pm \sqrt{4|\rho_{eg} (t)|^2 +(\rho_{ee} (t)-\rho_{gg}(t))^2}\Big)$ are the eigenvalues of $\rho_a (t)$.

Interestingly, from Eqs. (\ref{l1normcoherence}) and (\ref{populationinversion}), we find $\lambda_{\pm} = \tfrac{1}{2}\Big(1 \pm \sqrt{(C_{l_1}(t))^2 +(W(t))^2}\Big)$. Consequently, Eq. \eqref{vonNeumannJC} becomes
\begin{align}\label{vonNeumannEvolved}
S= 1 -\frac{1}{2} \Big(1 + \sqrt{(C_{l_1}(t))^2 +(W(t))^2}\Big) \log_2 \left[1 + \sqrt{(C_{l_1}(t))^2 +(W(t))^2} \right] \nonumber\\
- \frac{1}{2} \Big(1 - \sqrt{(C_{l_1}(t))^2 +(W(t))^2}\Big) \log_2 \left[ 1 - \sqrt{(C_{l_1}(t))^2 +(W(t))^2}\right] .
\end{align}
This expression connects the dynamics of atomic observables, which are the properties of a single-partite system, to atom-field entanglement, a property of a bipartite system. Such connections between the properties of single-partite and bipartite systems have been studied for qubit \cite{Jakob2010}  and qudit systems \cite{Jakob2007}.

It is important to note that Eq. \eqref{vonNeumannEvolved} is true for any initial pure atomic states and pure field states for both resonant and non-resonant cases.  %Here, $C_{l_1}$ is the $l_1$-norm of coherence \cite{Baumgratz2014Quantifying}, defined as
%\begin{align}
%C_{l_1}=2\,|\rho_{eg}|,
%\end{align}
%which is a basis-dependent measure of coherence. In a chosen basis, here $\{\ket{e},\ket{g}\}$, the off-diagonal elements of a density matrix represent the coherence between $\ket{e}$ and $\ket{g}$, which is quantified by $C_{l_1}$. The atomic population inversion  \cite{Tsutsui2025}
%\begin{align}
%W=\rho_{ee}-\rho_{gg},
%\end{align}
%is the difference in the probability of finding the atom in the excited state to the probability of finding the atom in the ground state. The value of $W$ changes when the exchanges energy with the field.
Therefore, the emergence of entanglement can be understood in terms of both atomic coherence $(C_{l_1}(t))$ and atomic population inversion $(W(t))$. Specifically, the quantity $(C_{l_1}(t))^2 +(W(t))^2$, which satisfies $0\leq (C_{l_1}(t))^2 +(W(t))^2\leq 1$, governs the degree of entanglement. The entanglement in the state $|\Psi_{af}(t) \rangle$ is zero if $(C_{l_1}(t))^2 +(W(t))^2=1$, whereas the entanglement is maximum for $(C_{l_1}(t))^2 +(W(t))^2=0$. Therefore, the atom-field system becomes maximally entangled when both $C_{l_1}(t)$ and $W(t)$ become zero simultaneously:
\begin{align}
    W(t)=C_{l_1}(t)=0 \Rightarrow S(t)=1.
\end{align}

The above results can also be understood geometrically in terms of the length of the Bloch vector on the Bloch sphere. We find that the magnitude of the Bloch vector that represents the state $\rho_{a} (t)$ is $|\vec{r}|=\sqrt{(C_{l_1}(t))^2 +(W(t))^2}$ (see \ref{AppendixBlochVector}). Therefore,  Eq. \eqref{vonNeumannEvolved} becomes
\begin{align}
    S(t) = 1-\frac{1}{2} \Big[ \big(1+|\vec{r}|\big) \log_2 \big(1+|\vec{r}|\big) + \big(1-|\vec{r}|\big) \log_2 \big(1-|\vec{r}|\big) \big],
\end{align}
which leads to maximum entanglement when $|\vec{r}|=0$, corresponding to a maximally mixed atomic state $\rho_a (t)$.

\section{Persistence of Maximal Atom–Field Entanglement}
To understand the persistence of maximal atom-field entanglement, we study the interplay between atomic coherence $(C_{l_1}(t))$ and atomic population inversion $(W(t))$ by considering various initial field states. For simplicity, we consider the initial state of the atom to be $\ket{e}$, so that the element $\rho_{ee}=1$, and all other elements are zero in the atomic density matrix $\rho_a(0)$. Therefore, at time $t=0$, we have $C_{l_1} (0)=0$ (no coherence between $\ket{e}$ and $\ket{g}$) and the population inversion $W(0)=1$. With this initial condition, we identify two distinct types of dynamical behavior: (a) the atomic coherence remains zero throughout the evolution, and (b) the atomic coherence develops during the evolution.  Based on these temporal behaviors, we investigate the conditions under which the maximal atom–field entanglement becomes persistent or transient.

%A \(2 \times 2\) density matrix, corresponding to a two-level system, can be written as 
%\begin{align*}
%\rho_a =\begin{bmatrix}\rho_{11} & \rho_{12} \\ \rho_{21} & \rho_{22}
%\end{bmatrix}, 
%\end{align*}
%\begin{align}\label{coherence}
%\rho_a= \tfrac{1}{2}\Big(I + \langle \sigma_x \rangle \sigma_x + \langle \sigma_y \rangle \sigma_y + \langle \sigma_z \rangle \sigma_z \Big),
%\end{align}
%where $\rho_{22}=1-\rho_{11}$,\quad $\rho_{21}=\rho_{12}^*$.
%\\A widely used measure of coherence we considered here is the $l_1$-norm of coherence, which is defined as 
%\begin{align}
%C_{l_1}(\rho_a)=\sum_{i \neq j}|\rho_{ij}|.
%\end{align}
%This is a basis dependent measure of coherence. In a particular basis, here we consider atomic basis $\{ \ket{e}, \ket{g} \}$, if the off-diagonal elements of $\rho_a$ are zero, then the state $\rho_a$ has zero coherence. Otherwise, $\rho_a$ has non-zero coherence.
%\\From Eq.(\ref{coherence}), we get 
%\begin{align}
%C_{l_1}(\rho_a)=\sqrt{\langle \sigma_x \rangle^2 +\langle \sigma_y \rangle^2}.
%\end{align}

\subsection{Vanishing atomic coherence}
We identify a class of initial field states for which the $l_1$-norm of coherence of the atom remains zero throughout the evolution, i.e., $C_{l_1} (t)=0$ for all $t$. As follows from Eq. \eqref{atomiccoherence} that this occurs if and only if either $C_n=0$ or $C_{n+1}=0$ (for every $n$) in an initial field state $\ket{\psi_f}$. Equivalently, the field state must have a photon-number distribution with no two adjacent occupied photon-number levels. Several well-known nonclassical field states, including Fock states, even coherent states, odd coherent states, squeezed vacuum states,  squeezed number states, photon-added squeezed vacuum states, photon-subtracted squeezed vacuum states, and compass states satisfy this condition. The corresponding Fock state representations of these states are given in \ref{AppendixCalculation}. Interestingly, the photon-number distributions of these field states have a special structure: their Fock-basis expansion contains photon-number states which are separated by fixed integer steps. Such field states are invariant under discrete phase space rotations (see \ref{DRS}) \cite{grimsmo2020quantum, lopez2020general}. However, in general, discrete rotational symmetry alone cannot fully be the reason for the persistent vanishing of the local atomic coherence during the evolution. Rather, the absence of adjacent occupied photon-number levels ensures that the atomic coherence remains zero at all times. 

As the $l_1$-norm of coherence remains zero throughout the evolution, the atom-field entanglement is determined solely by the atomic population inversion $W(t)$. Therefore, Eq. \eqref{vonNeumannEvolved} becomes
\begin{align}\label{entropy}
S(t) = 1-\frac{1}{2} \Big[ \big(1+W(t)\big) \log_2 \big(1+W(t)\big) + \big(1-W(t)\big) \log_2 \big(1-W(t)\big) \big].
\end{align}
The atom and the field are maximally entangled $(S=1)$ when $W(t)=0$. Therefore, for any initial field states for which the atomic population inversion exhibits complete collapse $(W(t)\approx 0)$ for a longer period of time, the atom-field entanglement persists at its maximum over the same period of time.

%The atomic population inversions, when the field is initialized in various field states, are
%\begin{align}
%W(t) &= \cos\!\big(2 {\Omega}_{n}\,t\big),~~~~(\text{Fock state})\\
%W(t) &= \frac{2e^{-|\alpha|^2}}{\left(1+e^{-2|\alpha|^2}\right)} \sum_{n\, even}^{\infty} \frac{|\alpha|^{2n}}{n!}\, \cos 2\Omega_n t,~~~~(\text{Even coherent state})\\
%W(t) &= \frac{2e^{-|\alpha|^2}}{\left(1-e^{-2|\alpha|^2}\right)} \sum_{n\, odd}^{\infty} \frac{|\alpha|^{2n}}{n!}\cos(2\Omega_n t), ~~~~(\text{Odd coherent state})\\
%W(t) &=\frac{1}{\cosh r} \,\sum_{n=0}^{\infty} \frac{{(2n)!}}{(2^n n!)^2} \left( \tanh r \right)^{2n}\,\cos\!\big(2 {\Omega}_{2n}\,t\big), ~~~~(\text{Squeezed vacuum})
%\end{align}
%\textcolor{red}{Need to add the population inversions for squeezed number states, photon-added squeezed vacuum, photon-subtracted squeezed vacuum.}
\begin{table}[h]
\small
\centering
\begin{tabular}{|p{5cm}|p{4cm}|p{4cm}|p{2.5cm}|}
%\begin{tabular}{|c|c|c|c|}
\hline
Initial field states & $l_1$-norm of coherence of atom & Atomic population inversion & Entanglement \\ \hline\hline
Fock states \cite{gerry2023introductory}  & Vanishing coherence & Does not show collapse  & Transient \\ \hline
Even coherent states \cite{dodonov1974even}  & Vanishing coherence  & Shows collapse  & Persistent \\ \hline
Odd coherent states \cite{dodonov1974even}  & Vanishing coherence & Shows collapse & Persistent \\ \hline
Squeezed vacuum \cite{stoler1970equivalence} & Vanishing coherence & Does not show collapse & Transient \\ \hline
Squeezed number states \cite{kim1989properties} & Vanishing coherence & Does not show collapse & Transient \\ \hline
Photon-added squeezed vacuum \cite{kun2010nonclassicality} & Vanishing coherence & Shows collapse & Persistent \\ \hline
Photon-subtracted squeezed vacuum \cite{meng2012nonclassicality} & Vanishing coherence & Shows collapse & Persistent \\ \hline
Compass states \cite{zurek2001sub,agarwal2004mesoscopic} & Vanishing coherence & Shows collapse & Persistent \\ \hline
Coherent states \cite{glauber1963coherent} & Dynamically generated coherence & Shows collapse & Transient \\ \hline
Yurke-Stoler states \cite{yurke1986generating}  & Dynamically generated coherence & Shows collapse & Persistent \\ \hline
Photon added coherent states \cite{PhysRevA.43.492}  & Dynamically generated coherence & Shows collapse & Transient \\ \hline
Displaced number states \cite{de1990properties}  & Dynamically generated coherence & Shows collapse & Transient \\ \hline
Squeezed coherent state \cite{stoler1970equivalence} & Dynamically generated coherence & Shows collapse & Transient \\ \hline
Binomial state \cite{stoler1985binomial} & Dynamically generated coherence & Shows collapse & Transient \\ \hline
Single number-state filtered coherent state \cite{Meher2018} & Dynamically generated coherence & Does not show collapse & Transient \\ \hline
\end{tabular}
\caption{Table for temporal behavior of atomic coherence, atomic population inversion, and entanglement for various initial field states}\label{mytable}
\end{table}
We found that, among the aforementioned states, only even coherent states, odd coherent states, photon-added squeezed vacuum states, photon-subtracted squeezed vacuum states, and compass states, with larger mean photon numbers, exhibit complete collapse $(W(t)\approx 0)$ for a longer period when they interact resonantly with the atom, and hence, they generate maximal atom-field entanglement that persists over an extended period. The minimum mean photon number required for the aforementioned field states to exhibit complete collapse is given in Table \ref{TableWcollapse} in \ref{AppendixCollapse}. For any arbitrary initial field state, the collapse of atomic population inversion is determined by its photon-number distribution, as can be seen from Eq. \eqref{populationinversion}. An appropriate photon-number distribution can lead to destructive interference among Rabi oscillations associated with individual photon-number states. Therefore, the persistence of maximal atom-field entanglement also depends on how the photon-number distribution spreads over the Fock space.

\begin{figure}[h]
    \centering
    \includegraphics[width=0.8\linewidth]{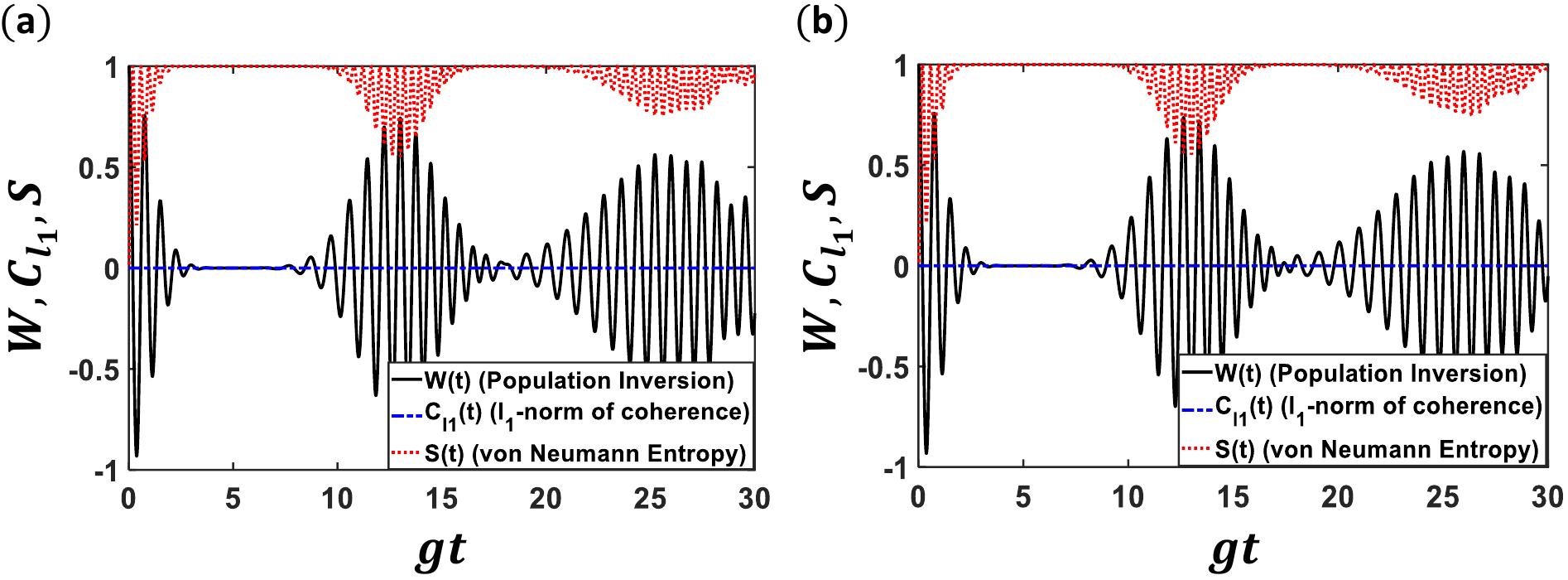}
    \caption{Atomic population inversion (black-continuous), $l_1$-norm of coherence (blue-dash-dot), and entanglement (red-dot) as a function of normalized time $gt$ when the cavity field is prepared in an (a) even coherent state, and (b) odd coherent state with $\alpha=4$.}
    \label{ECS}
\end{figure}

We show the atom-field entanglement $S(t)$, the $l_1$-norm of coherence $C_{l_1}(t)$, and the atomic population inversion $W(t)$ as a function of normalized time $gt$ in Fig. \ref{ECS} when the field is initialized in an (a) even coherent state and (b) odd coherent state. For these initial states, we calculate $W(t)$, $C_{l_1}(t)$, and $S(t)$ using Eqs. \eqref{populationinversion}, \eqref{atomiccoherence}  and \eqref{vonNeumannEvolved}, respectively. As the atom displays complete collapse $(W(t)\approx 0)$ over an extended period of time when it interacts with these initial field states, the atom-field entanglement attains its maximum value $(S=1)$ and persists for the same period of time. Although we have discussed the dynamics specifically for even and odd coherent states, we observed similar behavior for photon-added squeezed vacuum states, photon-subtracted squeezed vacuum states, and compass states.

In contrast, under non-resonant conditions, $W(t)$ approaches a nonzero constant value for an extended period of time (see Eq. \eqref{Wnonresonant} in \ref{AppendixNonresonant}). As a result, the entanglement persists at a value less than unity in that period (see Fig. \ref{Nonresonant}). Nevertheless, the entanglement is transient when $W(t) \approx 0$ and $C_{l_1} (t)=0$, and reaches its maximum value (see Fig. \ref{Nonresonant}).

On the other hand, states such as Fock states, squeezed vacuum states, and squeezed number states do not display complete collapse, even though their photon-number distributions spread across multiple photon numbers (except for Fock states). Consequently, they do not generate persistent maximal entanglement; instead, the maximal entanglement is transient (Table \ref{mytable}). The entanglement is maximum at the zero crossings of atomic population inversion. For a Fock state $\ket{N}$, the zero-crossing occurs at $t_m=(2m+1)\pi/(4g\sqrt{N+1})$ (see \ref{AppendixZeroCrossing}) and the interval between consecutive zero crossings decreases as the photon number $(N)$ increases. However, for the squeezed vacuum states and squeezed number states,  $W(t)$ does not exhibit regularly spaced zero crossings (see \ref{AppendixZeroCrossing}). Nevertheless, the time of the first zero-crossing of atomic population inversion decreases with increasing mean photon number of the field state (see Fig. \ref{SNS}).

%For even coherent states, we find (see Appendix)
%\begin{align} 
%C_{l_1}(t) =0, \quad W(t) = \frac{2e^{-|\alpha|^2}}{\left(1+e^{-2|\alpha|^2}\right)} \sum_{n\, even}^{\infty} \frac{|\alpha|^{2n}}{n!}\, \cos 2\Omega_n t,
%\end{align}
%and for odd coherent states
%\begin{align} 
%C_{l_1}(t) =0, \quad W(t) = \frac{2e^{-|\alpha|^2}}{\left(1-e^{-2|\alpha|^2}\right)} \sum_{n\, odd}^{\infty} \frac{|\alpha|^{2n}}{n!}\cos(2\Omega_n t).
%\end{align}

%Below, we discuss the entanglement dynamics for the number states and even coherent states in detail, while the corresponding discussion for the remaining states is presented in Appendix.

%This class includes well-known field states, including Fock states, even coherent states, odd coherent states, squeezed states, and thermal states. Below, we discuss the entanglement dynamics for the number states and even coherent states in detail, while the corresponding discussion for the remaining states is presented in Appendix.  

%The population inversion does not exhibit perfect collapse ($W(t)=0$ for a longer period) when it interacts with a number state. However, $W(t)=0$ at $t=(2k+1)\pi/4\Omega_n$, where $k$ is an integer. Therefore, the atom-field entanglement reaches to maximum at the discrete instants $t=(2k+1)\pi/4\Omega_n$ for which the population inversion vanishes, rather than over an extended interval.  

\subsection{Dynamically generated atomic coherence}
When an atom with zero coherence $(C_{l_1}=0)$ interacts with this class of field states, it gradually develops local atomic coherence. This class includes coherent states, Yurke-Stoler states, photon-added coherent states, displaced number states, squeezed coherent states, binomial states, and number-state filtered coherent states. These states do not satisfy the condition that no two adjacent photon-number states are populated in their Fock-basis expansions. Therefore, these states are not invariant under discrete phase space rotations. For these initial field states, the atom-field entanglement is determined by $W(t)$ and $C_{l_1}(t)$. Consequently, persistence of maximal atom-field entanglement becomes unlikely, since both $W(t)$ and $C_{l_1}(t)$ must simultaneously vanish for a longer period. Therefore, in most of the cases, the maximal atom-field entanglement is transient. The temporal behaviors of atomic coherence, atomic population inversion, and entanglement for these field states are tabulated in Table. \ref{mytable}. 

\begin{figure}[h]
    \centering
    \includegraphics[width=0.8\linewidth]{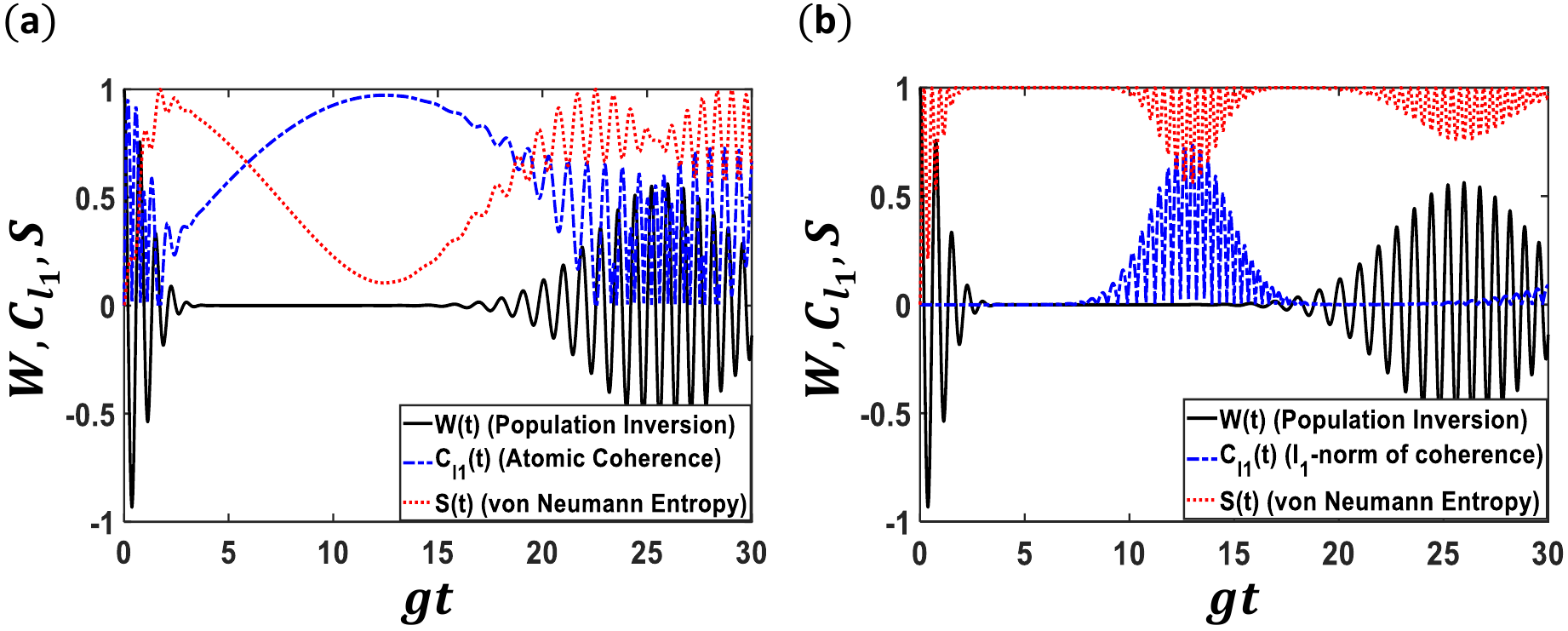}
    \caption{Atomic population inversion (black-continuous), $l_1$-norm of coherence (blue-dash-dot), and entanglement (red-dot) as a function of normalized time $gt$ when the cavity field is prepared in a (a) coherent state, and (b) Yurke-Stoler state with $\alpha=4$.}
    \label{Coherent}
\end{figure}

We show the atom-field entanglement $S(t)$, $l_1$-norm of coherence $C_{l_1}(t)$, and atomic population inversion $W(t)$ as a function of normalized time $gt$ in Fig. \ref{Coherent} when the field is initialized in a (a) coherent state and (b) Yurke-Stoler state. As can be seen in Fig. \ref{Coherent}(a), the $l_1$-norm of coherence $C_{l_1}(t)$ does not become zero when the field is in a coherent state. Therefore, the atom-field entanglement $S(t)$ becomes transient even though the atomic population inversion $W(t)$ displays collapse. This happens for other states too, except for the Yurke-Stoler states. For Yurke-Stoler states with a larger mean photon number, we find that the entanglement persists for a longer period (see Fig. \ref{Coherent}(b)). This is because, for Yurke-Stoler states, the $l_1$-norm of coherence grows very slowly from zero (see Fig. \ref{YS}) and remains much smaller than unity for an extended period (see Fig. \ref{Coherent}(b)). It is non-zero due to the constructive interference between two coherent state components $\ket{\alpha}$ and $\ket{-\alpha}$, which have a relative phase of $\pi/2$ (see \ref{AppendixYS}). This is also supported by the fact that the Yurke-Stoler state is not invariant under a discrete phase space rotation. However, there is a significant overlap between the region where $C_{l_1}(t)\ll 1$ and the atomic population inversion collapse region ($W(t)\approx 0)$. Consequently, the atom-field entanglement becomes nearly persistent in the overlapping region, as can be seen in Fig. \ref{Coherent}(b).

%We show the atom-field entanglement $S(t)$, the $l_1$-norm of coherence $C_{l_1}(t)$, and the atomic population inversion $W(t)$ as a function of normalized time $gt$ in Fig. \ref{Coherent}. As the $l_1$-norm of coherence is non-zero, the atom-field entanglement does not attain its maximum value for a longer period even though there is a collapse region $(W=0)$ for an extended period of time.   

%We show the atom-field entanglement $S(t)$, the $l_1$-norm of coherence $C_{l_1}(t)$, and the atomic population inversion $W(t)$ as a function of normalized time $gt$ in Fig. \ref{Yurke}. It is interesting to note that the $l_1$-norm of coherence remains zero initially and then revives. For an extended period of time, both the $l_1$-norm of coherence and the atomic-population inversion vanish; as a result, the atom-field entanglement reaches to its maximum value.  

\section{Summary}
We investigated the emergence and persistence of maximal atom–field entanglement in an atom-cavity system. By quantifying the atom-field entanglement using the von Neumann entropy, we found a direct relation between entanglement and the atomic dynamical variables such as the atomic coherence ($l_1$-norm of coherence), and atomic population inversion. This relation showed that the emergence of maximal atom–field entanglement is possible when the atomic population inversion and the $l_1$-norm of coherence simultaneously vanish.  Further, we considered the atom in its excited state, for which the $l_1$-norm of coherence of the atom is zero, and studied the role of initial field states on the time dynamics of entanglement. With this consideration, we showed that the persistence of maximal entanglement for a longer period depends strongly on the initial state of the field. We categorized the atom-field dynamics into two:\\
\noindent
(1) For initial field states, having photon-number distributions with no two consecutive photon-number levels occupied, the $l_1$-norm of coherence of the atom remains zero throughout the evolution. This condition is satisfied by well-known nonclassical states such as Fock states, even coherent states, odd coherent states, squeezed vacuum states,  squeezed number states, photon-added squeezed vacuum states, photon-subtracted squeezed vacuum states, and compass states.  Therefore, for these initial field states, the complete collapse of the atomic population inversion with zero local atomic coherence directly leads to maximal atom–field entanglement, which persists over an extended period.\\
\noindent
(2) For other initial field states, in which at least two consecutive photon-number levels are occupied, the local atomic coherence develops during the evolution, and the $l_1$-norm of coherence becomes non-zero. This condition is satisfied by states such as coherent states, Yurke-Stoler states, photon-added coherent states, displaced number states, squeezed coherent states, binomial states, and number-state filtered coherent states.  As a result, even when the population inversion vanishes, maximal entanglement is attained only for a short duration. Therefore, the atom-field entanglement shows transient behavior except for the Yurke-Stoler states. The Yurke-Stoler states with larger mean photon numbers exhibit persistent maximal atom-field entanglement.   

In summary, we demonstrated how the choice of the initial field state controls the temporal characteristics of entanglement in the Jaynes–Cummings system. The generation and persistence of maximal atom–field entanglement in such systems could be valuable for applications in quantum information processing. 

An interesting direction for future work is to extend the present study to more complex atom-field systems including the nonlinear Jaynes-Cummings models \cite{Manko1997,Uhdre2022, MedinaDozal2024}, where the nonlinear coupling between the atom and the field may influence the nature of atom-field entanglement for a given field state.

\appendix

\section{Quantum field states in the Fock basis}\label{AppendixCalculation}
%Here, we provide detailed calculations of the $l_1$-norm of coherence and the atomic population inversion for various initial field states when the atom is in its excited state.

%Consider the field state  $\ket{\psi_{{f}}(0)} = \sum_{n=0}^{\infty} C_n \ket{n}$ at time $t=0$, where $C_n$ are the probability amplitudes for the $n$-photon state. Under the evolution governed by the Hamiltonian $H$ given in Eq. \eqref{HJC}, the evolved state of the atom-field system at time $t$ is \cite{Meher2018,Sugumar2026,gerry2023introductory}
%\begin{align}\label{Appendixwft}
%\ket{\Psi_{af}(t)} = \sum_{n=0}^{\infty} C_n e^{-i\omega t}\left[ \cos\!\big({\Omega}_n\,t\big)\ket{e,n} - i \sin\!\big({\Omega}_n\,t\big)\ket{g,n+1}\right], 
%\end{align}
%where ${\Omega}_n=g\sqrt{n+1}$ is the Rabi frequency. 

%Now, we can calculate the $l_1$-norm of coherence and the atomic population inversion using
%\begin{align}\label{Appendixatomiccoherence}
%C_{l_1}=2\left|\sum_{n=0}^{\infty} C_{n+1}^* C_n 
%\cos\!\big({\Omega}_{n+1}\,t\big)\,\sin\!\big({\Omega}_n\,t\big)\right|.
%\end{align}
%and 
%\begin{align}\label{Appendixpopulationinversion}
% W(t)=\langle \sigma_z\rangle =\sum_{n=0}^\infty P_n \cos\left(2\Omega_n t\right),
%\end{align}
%where $P_n=|C_n|^2$ is the probability of finding $n$ photons in the field state. 

%As can seen from Eq. \eqref{Appendixwft}, one needs to know the probability amplitude $C_n$ for a given initial field state. Below, we have listed out the probability amplitudes for all the initial states that we have considered in the main text.
Here, we give the Fock state representation of various field states. In Fock basis $\{\ket{n}\}$, 
\begin{subequations}
\begin{itemize}
\item Coherent state \cite{glauber1963coherent} 
\begin{align}
\ket{\alpha} =e^{-|\alpha|^2/2} \sum_{n=0}^{\infty}\frac{\alpha^n}{\sqrt{n!}} \,\ket{n},
\end{align}
 where $\alpha$ is the coherent state amplitude.
 \item Even coherent state\, \cite{dodonov1974even}
\begin{align}
\ket{\psi_e}= \frac{1}{\sqrt{N_e}} \left(\ket{\alpha}+\ket{-\alpha} \right) =\frac{e^{-|\alpha|^2/2}}{\sqrt{2\left(1+e^{-2|\alpha|^2}\right)}} \sum_{n=0}^{\infty} \frac{\alpha^n}{\sqrt{n!}}(1+(-1)^n)\,\ket{n},
\end{align}
where $N_e=2\left(1+e^{-2|\alpha|^2}\right)$.
\item Odd coherent state\, \cite{dodonov1974even}
\begin{align}
\ket{\psi_o}=\frac{1}{\sqrt{N_o}} \left(\ket{\alpha}-\ket{-\alpha} \right)=\frac{e^{-|\alpha|^2/2}}{\sqrt{2\left(1-e^{-2|\alpha|^2}\right)}} \sum_{n=0}^{\infty}\frac{\alpha^n}{\sqrt{n!}}(1-(-1)^n)\,\ket{n},
\end{align}
where $N_o=2\left(1-e^{-2|\alpha|^2}\right)$.
\item Yurke-Stoler state \,\cite{yurke1986generating}
\begin{align}\label{YS}
\ket{\psi_{YS}}=\frac{1}{\sqrt{2}} \left(\ket{\alpha}+ i\,\ket{-\alpha} \right)=\frac{e^{-|\alpha|^2/2}}{\sqrt{2}} \sum_{n=0}^{\infty}\frac{\alpha^n}{\sqrt{n!}}(1+i(-1)^n)\,\ket{n}.
\end{align} 
\item Squeezed vacuum state\, \cite{gerry2023introductory}
\begin{align}
\ket{\xi} = \frac{1}{\sqrt{\cosh r}} \,\sum_{n=0}^{\infty} \frac{\sqrt{(2n)!}}{2^n n!} 
\left(-e^{i\theta} \tanh r \right)^n\,\ket{2n},
\end{align}
where $\xi=re^{i\theta}$ is the squeezing parameter, $r$ is the squeezing strength, and $\theta$ is the squeezing angle.
\item $m$-Photon added coherent state \, \cite{PhysRevA.43.492}
\begin{align}
\ket{\alpha, m} = \frac{e^{-|\alpha|^2/2}}{\sqrt{L_m(-|\alpha|^2) m!}} \sum_{n=0}^{\infty}\frac{\alpha^n}{n!}\sqrt{(n+m)!}\,\ket{n+m},
\end{align}
where $m$ is the number of photons added to the coherent state $\ket{\alpha}$ and $L_m(-|\alpha|^2)$ is the  Laguerre polynomial of order $m$.
\item Displaced number state \, \cite{de1990properties}
\begin{align}
\ket{\alpha, n}= e^{-|\alpha|^2/2} \, \sum_{m=0}^{\infty} \sqrt{\frac{n!}{m!}} \, \alpha^{\,(m-n)} \,L_n^{(m-n)}(|\alpha|^2)\ket{m},
\end{align}
where $L_n^{(m-n)}$ is the associated Laguerre polynomial of order $n$ with parameter $m-n$.
\item Number state filtered coherent state\, \cite{Meher2018}
\begin{align}
\ket{\psi (\alpha, m)} = \frac{e^{-|\alpha|^2/2}}{\sqrt{1-e^{-|\alpha|^2}\frac{|\alpha|^{2m}} {m!}}}\sum_{n=0,\, n \ne m}^{\infty} \frac{\alpha^n}{\sqrt{n!}}\,\ket{n}.
\end{align}
This state is formed by removing the number state $\ket{m}$ from the coherent state $\ket{\alpha}$.
\item Squeezed number state\, \cite{nieto1997displaced}
\begin{align}
S(\xi)|n\rangle = \frac{\sqrt{n!}}{\sqrt{N_{SNS}}(x)^{n+\frac{1}{2}}}
\sum_{j=0}^{ [\frac{n}{2}] }
\frac{(-d^*)^j (x)^{2j}}{(n-2j)! \, j!}
\sum_{k=0}^{\infty}
\frac{d^k}{k!}\sqrt{(n-2j+2k)!}\,|n-2j+2k\rangle,
\end{align}
where $N_{SNS}$ is the normalization constant, $d = \tfrac{1}{2} e^{i\phi}\tanh r$, $r$ and $\phi$ are squeezing parameters, $S(\xi)=\exp {\left[\frac{1}{2} (\xi^* a^2 - \xi a^{\dagger^2})\right]}$ is the squeeze operator, $ x= \cosh r$, and $[\frac{n}{2}]$ is the greatest integer function.
\item $m$-Photon added squeezed vacuum state\, \cite{kun2010nonclassicality}
\begin{align}
\ket{\xi, m}_{add}= \frac{1}{\sqrt{m!} \, (x)^{m+\tfrac{1}{2}}} \,
\sum_{n=0}^{\infty}\frac{\sqrt{(2n+m)!}}{2^n n!} \left(-e^{i\theta} \frac{y}{x} \right)^n\,\ket{2n+m},
\end{align}
where $m$ is the number of photons added to the squeezed vacuum state, and $y= \sinh r$ .
\item $m$-Photon subtracted squeezed vacuum state\, \cite{meng2012nonclassicality}
\begin{align}
\ket{\xi, m}_{sub}= \frac{1}{\sqrt{m!} \, (y)^m \sqrt{x}} \,\sum_{\left[n=\frac{m}{2}\right]}^{\infty} \frac{(2n)!}{2^n n! \sqrt{(2n-m)!}} \left(-e^{i\theta} \frac{y}{x} \right)^n\,\ket{2n-m},
\end{align}
where $m$ is the number of photons subtracted from the squeezed vacuum state.
\item Squeezed coherent state \, \cite{gerry2023introductory}
\begin{align}
\ket{\alpha, \xi} =\frac{1}{\sqrt{x}} \exp\!\left(-\tfrac{1}{2}|\alpha|^2 - \tfrac{1}{2} e^{i\theta}\alpha^{*2} \frac{y}{x}\right)\,
\sum_{n=0}^{\infty}\frac{1}{\sqrt{n!}} \left( \frac{e^{i\theta}\frac{y}{x}}{2} \right)^{n/2}H_n\!\left(\frac{\gamma}{\sqrt{e^{i\theta}2xy}}\right)\ket{n},
\end{align}
where $H_n\!\left(\frac{\gamma}{\sqrt{e^{i\theta}2xy}}\right)$ is the Hermite polynomial of order $n$ and $\gamma=\alpha\, x +\alpha^* e^{i\theta} y$.
\item Binomial state \, \cite{MANDAL2019193}
\begin{align}
\ket{p, M}=\sum_{n=0}^{M}\left[{}^{M}C_{n} \, p^{n} (1-p)^{M-n} \right]^{\tfrac{1}{2}}\ket{n},
\end{align}
where $M$ is the maximum photon number and $p$ is the probability parameter.
\item Compass state\,  \cite{zurek2001sub,agarwal2004mesoscopic}
\begin{align}
\ket{\psi_{compass}}=\frac{e^{-|\alpha|^2/4}}{\sqrt{2}} \sum_{n=0}^{\infty}\frac{(\frac{\alpha}{\sqrt{2}})^n}{\sqrt{n!}}(1+(-1)^n+i^n+(-i)^n)\ket{n}.
\end{align}
\end{itemize}
\end{subequations}

\section{Non-resonant dynamics}\label{AppendixNonresonant}
Under the non-resonance condition $\omega \ne \omega_0$, 
%the energy eigenvalues of the Hamiltonian $H$ are $E_{\pm} = \hbar \omega (n+1) \pm \tfrac{1}{2}\hbar\, \tilde{\Omega}_{n}$ with the corresponding eigenvectors
%$\ket{n_{+}} = \sqrt{\frac{\tilde{\Omega}_{n}+\Delta}{2 \tilde{\Omega}_{n}}} \ket{e,n} +  \sqrt{\frac{\tilde{\Omega}_{n}-\Delta}{2 \tilde{\Omega}_{n}}} \ket{g,n+1} $ and  $\ket{n_{-}} = - \sqrt{\frac{\tilde{\Omega}_{n}-Delta}{2 \tilde{\Omega}_{n}}} \ket{e,n} +  \sqrt{\frac{\tilde{\Omega}_{n}+\Delta}{2 \tilde{\Omega}_{n}}} \ket{g,n+1}$, where $\tilde{\Omega}_{n}=\sqrt{\Delta^2 + 4 g^2 (n+1)}$.
the evolved state of the atom-field system at time $t$ is \cite{larson2021jaynes} 
\begin{align}\label{dwft}
\ket{\Psi_{af}(t)} = \sum_{n=0}^{\infty} C_n e^{-i(n+1)\omega t}\left[ \left( \cos\left(\frac{\tilde{\Omega}_{n}\,t}{2}\right)- \frac{i \Delta}{\tilde{\Omega}_{n}}\, \sin\left(\frac{\tilde{\Omega}_{n}\,t}{2} \right) \right) \ket{e,n} \right.\nonumber\\
\left. - i\, \frac{2g \sqrt{n+1}}{\tilde{\Omega}_{n}}\sin\left(\frac{\tilde{\Omega}_{n}\,t}{2}\right)\ket{g,n+1}\right],
\end{align}
where $\tilde{\Omega}_{n}=\sqrt{\Delta^2 + 4 g^2 (n+1)}$ and $\Delta=\omega_0-\omega$. 

The density matrix corresponding to the atom-field state is $\rho_{af}(t) = \ket{\Psi_{af}(t)}\!\bra{\Psi_{af}(t)}$. By taking the partial trace over the field degree of freedom, we get the reduced density matrix for the atom whose matrix elements are
\begin{subequations}
\begin{align}
\rho_{ee}(t) &= \sum_{n=0}^{\infty} |C_n|^2 \left(\cos^2\left(\frac{\tilde{\Omega}_{n}\,t}{2}\right) + \frac{\Delta^2}{(\tilde{\Omega}_{n})^2}\, \sin^2\left(\frac{\tilde{\Omega}_{n}\,t}{2}\right) \right),\\
 \rho_{eg}(t) &= 2g\,i\,  e^{-i\omega t} \sum_{n=0}^{\infty} C_{n+1} C_n^* \frac{\sqrt{(n+1)}}{\tilde{\Omega}_{n}}\, \sin\left(\frac{\tilde{\Omega}_{n}\,t}{2}\right) \left( \cos\left(\frac{\tilde{\Omega}_{n+1}\,t}{2}\right)- \frac{i \Delta}{\Omega_{n+1}}\sin\left(\frac{\tilde{\Omega}_{n+1}\,t}{2}\right) \right),\\
\rho_{ge}(t) &= -2g\,i\, e^{i\omega t} \sum_{n=0}^{\infty} C_n C_{n+1}^*
\frac{\sqrt{(n+1)}}{\tilde{\Omega}_{n}}\, \sin\left(\frac{\tilde{\Omega}_{n}\,t}{2}\right) \left( \cos\left(\frac{\tilde{\Omega}_{n+1}\,t}{2}\right)+ \frac{i \Delta}{\Omega_{n+1}}\sin\left(\frac{\tilde{\Omega}_{n+1}\,t}{2}\right) \right),\\
\rho_{gg}(t) &= 4g^2\, \sum_{n=0}^{\infty} |C_n|^2 \, \frac{(n+1)}{(\tilde{\Omega}_{n})^2}\, \sin^2\left(\frac{\tilde{\Omega}_{n}\,t}{2}\right).
\end{align}
\end{subequations}

From these matrix elements, we calculate the $l_1$-norm of coherence of the atom to be
\begin{align}
C_{l_1}(t) = 2 \left| \sum_{n=0}^{\infty} C_{n+1} C_n^* 
\left( \frac{2g\sqrt{n+1}}{\tilde{\Omega}_n}\sin\!\left(\frac{\tilde{\Omega}_n t}{2}\right) \right) \left( \cos\!\left(\frac{\tilde{\Omega}_{n+1} t}{2}\right) - i \frac{\Delta}{\tilde{\Omega}_{n+1}}\sin\!\left(\frac{\tilde{\Omega}_{n+1} t}{2}\right) \right) \right|,
\end{align}
and the atomic population inversion to be
\begin{align}\label{Wnonresonant}
W(t) = \sum_{n=0}^{\infty} |C_n|^2  \frac{\Delta^2}{\tilde{\Omega}_n^2} + \sum_{n=0}^{\infty} |C_n|^2 \,\frac{4g^2(n+1)}{\tilde{\Omega}_n^2}\cos(\tilde{\Omega}_n t).
\end{align}

\begin{figure}[h]
    \centering
    \includegraphics[width=0.5\linewidth]{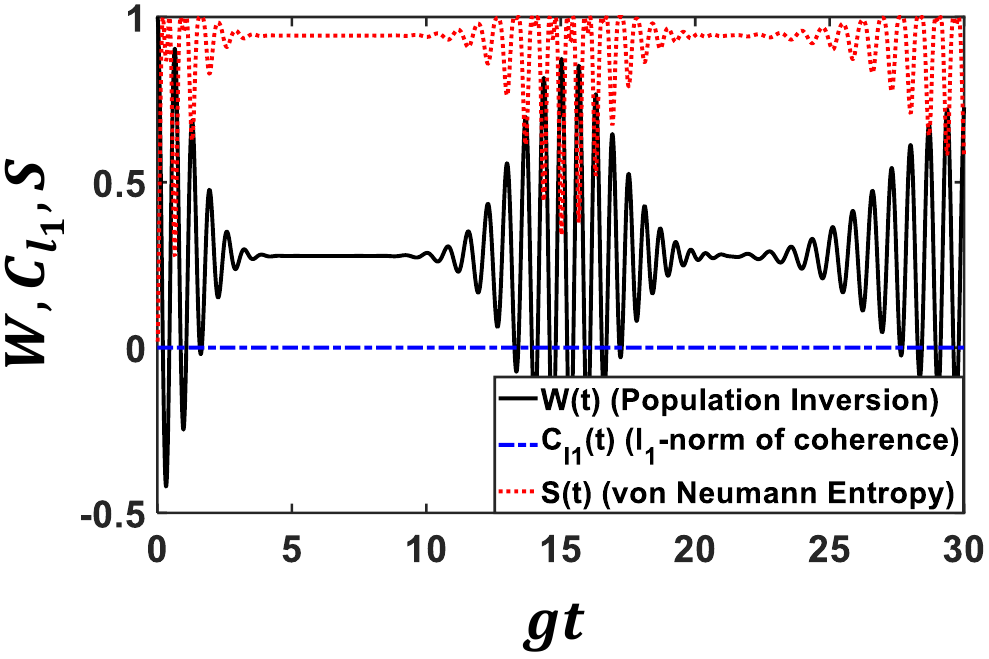}
    \caption{Atomic population inversion (black-continuous), $l_1$-norm of coherence (blue-dash-dot), and entanglement (red-dot) as a function of normalized time $gt$ for the non-resonant case with $\Delta=0.5$. The cavity field is prepared in an even coherent state with $\alpha=4$.}
    \label{Nonresonant}
\end{figure}

We show $S(t)$, $C_{l_1}(t)$, $W(t)$ as a function of normalized time $gt$ in Fig. \ref{Nonresonant} for the non-resonant case. The field is initialized in an even coherent state. As can be seen in the figure,  $C_{l_1}(t)$ remains zero in the non-resonant case too. However, as we note from  Eq. \eqref{Wnonresonant}, $W(t)$ contains an oscillatory term and a non-oscillatory term. For an extended period of time, the oscillatory term vanishes and $W(t)$ equalizes to a constant term 
\begin{align}
W(t) = \sum_{n=0}^{\infty} |C_n|^2  \frac{\Delta^2}{\tilde{\Omega}_n^2}.    
\end{align}
Consequently, the entanglement does not reach its maximum value $S=1$ in that duration, as can be seen in Fig. \ref{Nonresonant}.

\section{Bloch vector, $l_1$-norm of coherence and population inversion}\label{AppendixBlochVector}
A $2\times 2$ density matrix $(\rho_a)$ is represented by a vector on the Bloch sphere. The length of the Bloch vector is given by \cite{nielsen2010quantum}
\begin{align}
    |\vec{r}|&=\sqrt{\langle \sigma_x \rangle^2+\langle \sigma_y \rangle^2+\langle \sigma_z \rangle^2},
\end{align}
where $\sigma_x, \sigma_y$ and $\sigma_z$ are Pauli matrices and $\langle \sigma_s \rangle=\text{Tr}(\sigma_s \rho_a)$, with $s=x,y,z$. The length of the Bloch vector satisfies $(0\leq |\vec{r}|\leq 1)$. For pure states, $|\vec{r}|=1$ and for mixed states, $|\vec{r}|<1$.

Now, considering the reduced density matrix of the atom $\rho_a(t)$ given in Eq. \eqref{density matrix}, and using Eqs. \eqref{l1normcoherence}, and \eqref{populationinversion}, we get
\begin{align}
    |\vec{r}|&=\sqrt{4|\rho_{eg} (t)|^2 +(\rho_{ee} (t)-\rho_{gg}(t))^2}, \nonumber\\
    &=\sqrt{(C_{l_1}(t))^2 +(W(t))^2}.
\end{align}
Therefore,  Eq. \eqref{vonNeumannEvolved} becomes
\begin{align}
S(t) = 1-\frac{1}{2} \Big[ \big(1+|\vec{r}|\big) \log_2 \big(1+|\vec{r}|\big) + \big(1-|\vec{r}|\big) \log_2 \big(1-|\vec{r}|\big) \big].
\end{align}
The maximum atom-field entanglement occurs when $|\vec{r}|=0$, corresponding to the maximally mixed state of the atom.

\section{Discrete Rotational Symmetry}\label{DRS}
A field state $\ket{\psi_f}$ is said to have $k$-fold rotational symmetry in phase space if it is invariant under a phase-space rotation by an angle $2\pi/k$, up to a global phase \cite{grimsmo2020quantum, lopez2020general}. Mathematically,
\begin{align}
    {R}(\theta)\ket{\psi_f}=e^{i\phi}\ket{\psi_f},
\end{align}
where ${R}(\theta)=e^{-i\theta a^\dagger a}$ is the phase space rotation operator, $\theta=2\pi/k$, and $\phi$ is a phase. For density matrices,  
\begin{align}
R(\theta) \rho R^\dagger(\theta)=\rho.
\end{align}

As an example, to examine whether the even coherent state $\ket{\psi_e}$ is invariant under a discrete phase-space rotation, we consider the action of the rotation operator on it. We obtain,
\begin{equation}
\begin{aligned}
    {R}(\theta)\ket{\psi_e}
    &=\frac{e^{-|\alpha|^2/2}}{\sqrt{2\left(1+e^{-2|\alpha|^2}\right)}} \sum_{n=0}^{\infty} \frac{\alpha^n}{\sqrt{n!}}(1+(-1)^n)~e^{-i\theta a^\dagger a} \ket{n}, \nonumber \\
   &=\frac{e^{-|\alpha|^2/2}}{\sqrt{2\left(1+e^{-2|\alpha|^2}\right)}} \sum_{n=0}^{\infty} \frac{\alpha^n}{\sqrt{n!}}(1+(-1)^n)~e^{-i\theta n} \ket{n}.
\end{aligned}
\end{equation}
For $k=2$, we have $\theta=\pi$, then,
\begin{align}
    {R}(\theta)\ket{\psi_e}
    &=\frac{e^{-|\alpha|^2/2}}{\sqrt{2\left(1+e^{-2|\alpha|^2}\right)}} \sum_{n=0}^{\infty} \frac{\alpha^n}{\sqrt{n!}}(1+(-1)^n)~ (-1)^n \ket{n}, \nonumber \\
    &=\frac{e^{-|\alpha|^2/2}}{\sqrt{2\left(1+e^{-2|\alpha|^2}\right)}} \sum_{n=0}^{\infty} \frac{\alpha^n}{\sqrt{n!}}((-1)^n +1 ) \ket{n}, \nonumber \\ 
    &=\ket{\psi_e},
\end{align}
which shows that the even coherent state possesses a two-fold $(k=2)$ rotational symmetry. Similarly, we can show that odd coherent states, squeezed vacuum states, squeezed number states, photon‑added squeezed vacuum states, and photon‑subtracted squeezed vacuum states are also invariant under discrete phase space rotations with two-fold rotational symmetry. However, the compass states exhibit four-fold rotational symmetry. This rotational invariance of these states arises from the fact that they are superpositions of Fock states which are separated by fixed integer steps. Furthermore, a correspondence has been established between the rotational invariance of states and the states obtained after filtering Fock states separated by fixed integer steps \cite{lopez2020general}.

In contrast, states such as coherent states, Yurke–Stoler states, squeezed coherent states, photon‑added coherent states, displaced number states, and binomial states contain all Fock states. Therefore, they do not exhibit rotational invariance under a non-trivial discrete phase space rotation.

\section{Atomic population inversion}\label{AppendixCollapse}
The atomic population inversion for a field state having a photon-number distribution $P_n$ is \cite{gerry2023introductory}
\begin{align}\label{Appendixpopulationinversion}
 W(t)=\sum_{n=0}^\infty P_n \cos\left(2\Omega_n t\right).
\end{align}
For certain field states, $W(t)$ remains close to zero for an extended period, indicating the collapse of atomic population inversion. To quantify this behavior, we consider the following numerical criterion: collapse is considered to occur if $W(t)\leq10^{-4}$ persists for a duration of at least $g\Delta t \approx 1$. According to this criterion, the well-known field states listed in Table \ref{TableWcollapse} exhibit collapse when the mean photon number exceeds a state-dependent threshold value.

\begin{table}[h]
\small
\centering
%\begin{tabular}{|p{5cm}|p{4cm}|p{4cm}|p{2.5cm}|}
\begin{tabular}{|c|c|c|}
\hline
Initial field states & Minimum mean photon number \\ \hline \hline
Even coherent states  & $15$  \\ \hline
Odd coherent states & $14$  \\ \hline
Yurke-Stoler states & $7$  \\ \hline
Coherent states  & $7$ \\ \hline
Photon added squeezed vacuum & $17$  \\ \hline
Photon subtracted squeezed vacuum & $18$  \\ \hline
Compass states  &  $35$  \\ \hline
Photon added coherent states  & $5$ \\ \hline
Displaced number states  & $10$ \\ \hline
Squeezed coherent states  & $6$ \\ \hline
Binomial states  & $6$  \\ \hline
\end{tabular}
\caption{Minimum mean photon numbers required in various initial field states for the collapse of atomic population inversion in the Jaynes-Cummings system.}\label{TableWcollapse}
\end{table}

\section{Zero crossings of $W(t)$}\label{AppendixZeroCrossing}
For the Fock state $\ket{N}$, the probability amplitude is $C_n=\delta_{n,N}$. Therefore, from Eq. \eqref{populationinversion}, we calculate the atomic population inversion when the cavity field is in the Fock state $\ket{N}$ to be   
\begin{align}
 W(t)= \cos\left(2\Omega_N t\right).
\end{align}
Therefore, the zero-crossing occurs when $2\Omega_N t_m=\pi/2$, which gives 
\begin{align}
    t_m=\frac{(2m+1)\pi}{(4g\sqrt{N+1})}.
\end{align}

For the squeezed vacuum state, the atomic population inversion is
\begin{align}
W(t) = \frac{1}{\cosh r} \,\sum_{n=0}^{\infty} \frac{(2n)!}{(2^n n!)^2} 
\left(\tanh r \right)^{2n}\, \cos(2 \Omega_{2n} t).
\end{align}
For the squeezed-number state,
\begin{align}
W(t) = \frac{1}{{N_{SNS}}}\frac{n!}{(x)^{2n+1}}
\sum_{j=0}^{[\frac{n}{2}]}
\frac{|d|^{2j} (x)^{4j}}{[(n-2j)! \, j!]^2}
\sum_{k=0}^{\infty}
\frac{|d|^{2k}}{(k!)^2} \,(n-2j+2k)!\, \cos(2 \Omega_{n-2j+2k} t).
\end{align}

\begin{figure}[h]
    \centering
    \includegraphics[width=0.8\linewidth]{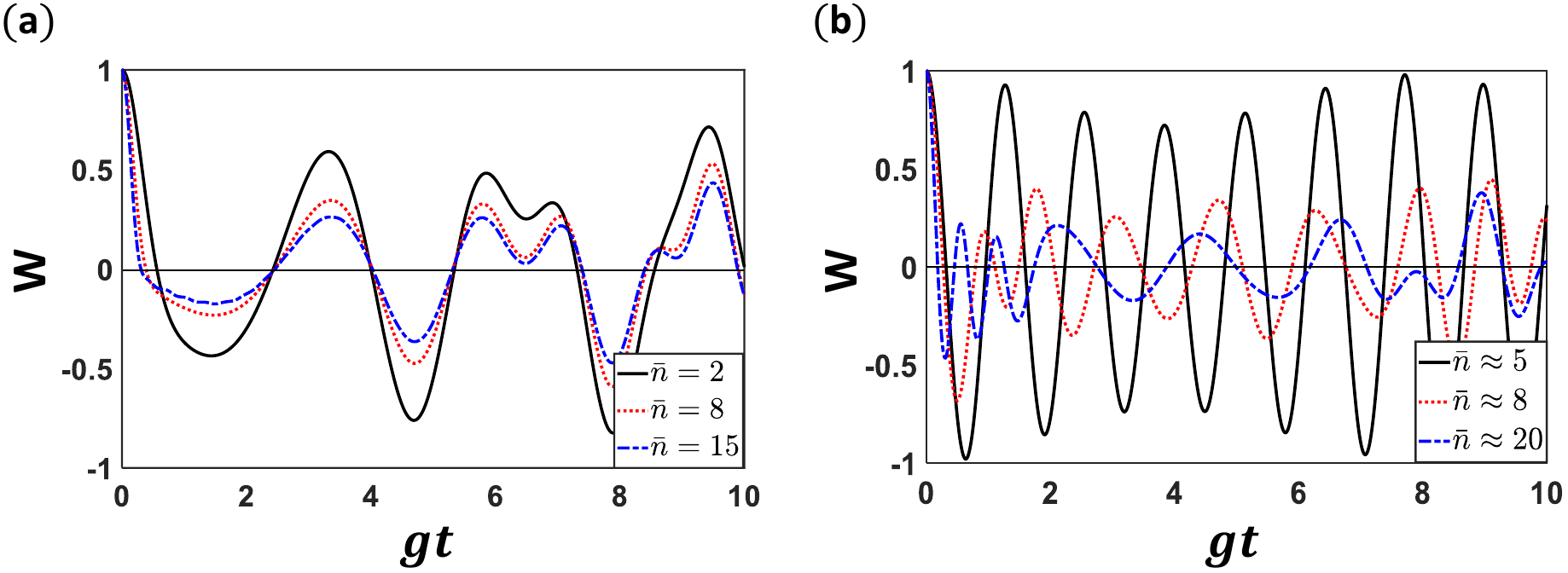}
    \caption{Atomic population inversion as a function of normalized time $gt$ when the cavity field is prepared in a (a) squeezed vacuum state with mean photon number $\bar{n}=2,8,$ and $15$ and (b) squeezed number state with mean photon number $\bar{n}=5,8,$ and $20$}.
    \label{SNS}
\end{figure}

In these cases, the atomic population inversion does not exhibit regular oscillatory behavior (see Fig. \ref{SNS}). As a result, a well‑defined period for the zero‑crossings can not be determined. Nevertheless, in both cases, the time of the first zero-crossing of the atomic population inversion decreases with increasing mean photon number of the field state (see Fig. \ref{SNS}). 

\section{Analysis for cat states}\label{AppendixYS}
Consider a superposition of two coherent states, $\ket{\alpha}$ and $\ket{-\alpha}$, with a relative phase $\phi$, given by

\begin{align} 
\ket{\psi_{cat}}= \frac{1}{\sqrt{N_c}} \left(\ket{\alpha}+e^{i \phi} \ket{-\alpha} \right),
\end{align}
where $N_c= 2(1+ e^{-2 |\alpha|^2} \cos{\phi})$. The $l_1$-norm of coherence of the above state is
\begin{align}\label{YSCl1}
C_{l_1}(t)
&=2\,\left|\sum_{n=0}^{\infty} C_{n+1} C_n^* \cos\!\big({\Omega}_{n+1}\,t\big)\,\sin\!\big({\Omega}_n\,t\big)\right|, \nonumber \\
&=\frac{2}{N_c}\, e^{-|\alpha|^2} \left|\sum_{n=0}^{\infty}\frac{|\alpha|^{n+1}}{\sqrt{(n+1)!}} \frac{\alpha^{*^n}}{\sqrt{n!}} \left(1+e^{i\phi}(-1)^{n+1} \right) \left(1+e^{-i\phi}(-1)^{n} \right) \cos\left({\Omega}_{n+1}\,t\right)\,\sin\left({\Omega}_n\,t\right)\right|, \nonumber \\
& = \frac{2}{N_c}\, |\alpha| \, e^{-|\alpha|^2} \, \left|\sum_{n=0}^{\infty}\frac{|\alpha|^{2n}}{n! \sqrt{(n+1)}}  \left(e^{i\phi}(-1)^{n+1} + e^{-i\phi}(-1)^{n} \right) \cos\left({\Omega}_{n+1}\,t\right)\,\sin\left({\Omega}_n\,t\right) \right|, \nonumber \\
& = \frac{2}{N_c}\, |\alpha| \, e^{-|\alpha|^2} \, \left|\sum_{n=0}^{\infty}\frac{|\alpha|^{2n}}{n! \sqrt{(n+1)}}  (-1)^{n+1} \left(e^{i\phi} - e^{-i\phi} \right) \cos\left({\Omega}_{n+1}\,t\right)\,\sin\left({\Omega}_n\,t\right) \right|,  \nonumber \\
& = \frac{2}{N_c}\, |\alpha| \, e^{-|\alpha|^2} \, \left|\sum_{n=0}^{\infty}\frac{|\alpha|^{2n}}{n! \sqrt{(n+1)}}  (-1)^{n+1} \left(2i \sin{\phi} \right) \cos\left({\Omega}_{n+1}\,t\right)\,\sin\left({\Omega}_n\,t\right) \right|, \nonumber\\
& = \frac{4}{N_c}\, |\alpha| \, e^{-|\alpha|^2} \sin\phi\, \left|\sum_{n=0}^{\infty}\frac{|\alpha|^{2n}}{n! \sqrt{(n+1)}}  (-1)^{n+1}  \cos\left({\Omega}_{n+1}\,t\right)\,\sin\left({\Omega}_n\,t\right) \right|.
\end{align}

As can be noted from the above equation, for $\phi=0$  corresponding to the even coherent state and $\phi=\pi$ corresponding to the odd coherent state, we get $C_{l_1} (t)=0$ for all $t$ due to the destructive interference between two coherent states $\ket{\alpha}$ and $\ket{-\alpha}$.

In the short‑time limit, we use $\cos(\Omega_{n+1}t)\approx 1$ and $\sin(\Omega_{n}t)\approx \Omega_{n}t$, and simplified the above expression to
\begin{align}
C_{l_1}(t) \approx \frac{4}{N_c}\, |\alpha|\, gt \, e^{-|\alpha|^2} \sin\phi \left|\sum_{n=0}^{\infty}\frac{|\alpha|^{2n}}{n!}  (-1)^{n+1}  \right|.
\end{align}

For $\phi=\pi/2$,  corresponding to the Yurke-Stoler state, the $l_1$-norm of coherence is approximately given by

\begin{align}\label{ApproximateYSCl1}
C_{l_1}(t)
& \approx 2\, |\alpha|\, gt \, e^{-|\alpha|^2} \left|\sum_{n=0}^{\infty}\frac{|\alpha|^{2n}}{n!}  (-1)^{n+1}  \right|,\nonumber \\
& \approx 2\, |\alpha| \, g\, t\, e^{-2|\alpha|^2},
\end{align}
which is non-zero, due to constructive interference between $\ket{\alpha}$ and $\ket{-\alpha}$. 
We compare this approximate result with the analytical expression given in Eq. \eqref{YSCl1} in Fig. \ref{YS}. This indicates that the $l_1$-norm of coherence is non-zero, although very small.

\begin{figure}[h]
    \centering
    \includegraphics[width=0.5\linewidth]{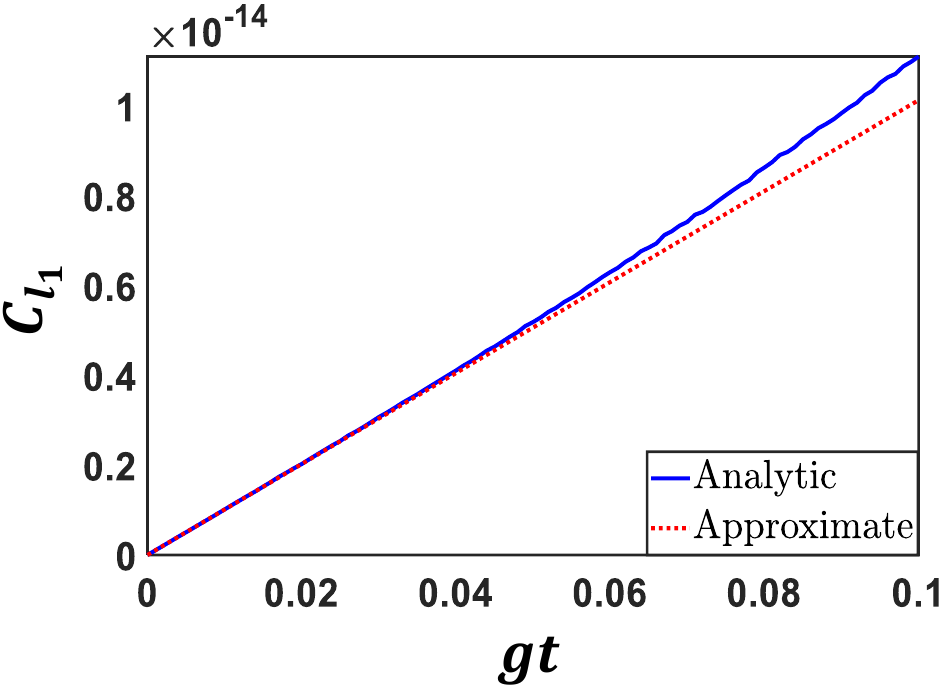}
    \caption{ Comparison between approximate result (Eq. \eqref{ApproximateYSCl1}) with analytical result (Eq. \eqref{YSCl1}) for $\alpha=3$.}
    \label{YS}
\end{figure}

%\section*{Data availability statement}
%No new data were created or analysed in this study. 
%\section*{Conflict of interest}
%The authors declare that they have no conflicts of interest.

%\bibliography{References}

\section*{Data availability statement}
No new data were created or analysed in this study. 
\section*{Conflict of interest}
The authors declare that they have no conflicts of interest.

\section*{References}
%\bibliographystyle{iopart-num}
%\bibliography{References}

\providecommand{\newblock}{}

\end{document}